\documentclass[apj]{emulateapj}

\shorttitle{Multi-Wavelength Constraints on the Day-Night Circulation Patterns of \mbox{HD~189733b}}
\shortauthors{Knutson et al.}\def\simgr{\,\hbox{\hbox{$ > $}\kern -0.8em \lower 1.0ex\hbox{$\sim$}}\,}
\def\simle{\,\hbox{\hbox{$ < $}\kern -0.8em \lower 1.0ex\hbox{$\sim$}}\,}

\begin{document}

\title{Multi-Wavelength Constraints on the Day-Night Circulation Patterns of HD~189733b}

\author{Heather A. Knutson\altaffilmark{1,2}, David Charbonneau\altaffilmark{1,3}, Nicolas B. Cowan\altaffilmark{4}, Jonathan J. Fortney\altaffilmark{5}, Adam P. Showman\altaffilmark{6}, Eric Agol\altaffilmark{4}, Gregory W. Henry\altaffilmark{7}, Mark E. Everett\altaffilmark{8}, and Lori E. Allen\altaffilmark{1}}
\altaffiltext{1}{Harvard-Smithsonian Center for Astrophysics, 60 Garden St., Cambridge, MA 02138}
\altaffiltext{2}{hknutson@cfa.harvard.edu}
\altaffiltext{3}{Alfred P. Sloan Research Fellow} 
\altaffiltext{4}{Department of Astronomy, Box 351580, University of Washington, Seattle, WA 98195}
\altaffiltext{5}{Department of Astronomy and Astrophysics, UCO/Lick Observatory, University of California, Santa Cruz, CA 95064}
\altaffiltext{6}{Lunar and Planetary Laboratory and Department of Planetary Sciences, University of Arizona, Tucson, AZ 85721}
\altaffiltext{7}{Center of Excellence in Information Systems, Tennessee State University, 3500 John A. Merritt Blvd., Box 9501, Nashville, TN 37209}
\altaffiltext{8}{Planetary Science Institute, 1700 E. Fort Lowell Rd., Suite 106, Tucson, AZ 85719}

\begin{abstract}

We present new \emph{Spitzer} observations of the phase variation of the hot Jupiter HD~189733b in the MIPS 24~\micron~bandpass, spanning the same part of the planet's orbit as our previous observations in the IRAC 8~\micron~bandpass \citep{knut07}.  We find that the minimum hemisphere-averaged flux from the planet in this bandpass is $76\pm3\%$ of the maximum flux; this corresponds to minimum and maximum hemisphere-averaged brightness temperatures of $984\pm48$~K and $1220\pm47$~K, respectively.  The planet reaches its maximum flux at an orbital phase of $0.396\pm0.022$, corresponding to a hot region shifted $20-30$\degr~east of the substellar point.  Because tidally locked hot Jupiters would have enormous day-night temperature differences in the absence of winds, the small amplitude of the observed phase variation indicates that the planet's atmosphere efficiently transports thermal energy from the day side to the night side at the 24~\micron~photosphere, leading to modest day-night temperature differences.  The similarities between the 8 and 24~\micron~phase curves for HD 189733b lead us to conclude that the circulation on this planet behaves in a fundamentally similar fashion across the range of pressures sensed by these two wavelengths.  One-dimensional radiative transfer models indicate that the 8~\micron~band should probe pressures $2-3$ times greater than at 24~\micron, although the uncertain methane abundance complicates the interpretation.  If these two bandpasses do probe different pressures, it would indicate that the temperature varies only weakly between the two sensed depths, and hence that the atmosphere is not convective at these altitudes.  We also present an analysis of the possible contribution of star spots to the time series at both 8 and 24~\micron~based on near-simultaneous ground-based observations and additional \emph{Spitzer} observations.  Accounting for the effects of these spots results in a slightly warmer night-side temperature for the planet in both bandpasses, but does not otherwise affect our conclusions. 

\end{abstract}


\section{Introduction}

We currently know of more than 30 transiting planetary systems, of which the majority are gas-giant planets orbiting extremely close ($<$0.05~A.U.) to their parent stars\footnote{See http://www.inscience.ch/transits for the latest count}.  These planets, known as ``hot Jupiters'', receive $>$10,000 times more radiation from their stars than Jupiter does from the Sun, heating them to temperatures as high as 2000~K \citep{har07}.  Most of these planets are expected to be tidally locked, with permanent day and night sides. As a result of this intense and highly asymmetric irradiation and their presumably slower rotation rates, the atmospheric dynamics of these planets are expected to differ significantly from those of the gas giant planets in the solar system.  

One of the fundamental questions regarding these planets is what fraction, if any, of the energy absorbed by the perpetually-illuminated day side is transferred to the night side.  The answer depends on the relative sizes of the radiative and advective time scales and may vary from planet to planet depending on the specific properties of the atmosphere.  Circulation models for these planets \citep[for a recent review see][]{show07} predict a range of possibilities, with day-night temperature differences ranging as high as $500-1000$~K \citep{show02,cho03,cho08,burkert05,coop05,coop06,lang07,dobb08,show08}.  The form of this circulation also varies, with some models predicting a quasi-steady state pattern consisting of one or several equatorial bands of winds circling the planet \citep{show02,burkert05,coop05,coop06,lang07,show08,dobb08}, and others predicting the formation of more complicated structures such as polar vortices whose positions may vary over time \citep{cho03,cho08}.

By observing the changes in the planet's thermal emission as a function of orbital phase, we can directly determine the day-night temperature difference for these hot Jupiters. \citet{har06} reported the first detection of these phase variations for the non-transiting planet $\upsilon$~Andromedae b at 24~\micron.  If one makes reasonable assumptions about the predicted size and temperature of $\upsilon$~Andromedae b based on its mass and distance from its star, the large size of the observed phase variation implies a large day-night temperature difference and correspondingly inefficient thermal homogenization between the day and night sides.  \citet{cow07} made similar 8~\micron~observations of three other systems, HD 209458, HD 179949, and 51 Peg, of which only HD 209458 is a transiting system.  They report a detection for the non-transiting system HD 179949, implying a large day-night temperature difference similar to that of $\upsilon$~Andromedae b.

Both the observations by \citet{har06} and \citet{cow07} are sparsely sampled, consisting of a series of brief visits with the \emph{Spitzer Space Telescope} spread out over the planet's orbit, and thus do not place any strong constraints on the timing of minima or maxima in the light curve.  The best-constrained determination of a phase curve for a hot Jupiter comes from \citet{knut07}, hereafter known as Paper I, where we present 8~\micron~observations of the transiting planet HD~189733b.  This planet has a mass of $1.14\pm0.06$~$M_{Jup}$ \citep{bouchy05,torr08} and a radius of $1.138\pm0.027$ $R_{Jup}$ \citep{bak06b,winn07a,knut07,pont07,pont08,torr08}, and orbits a K0V primary with a $V$-band magnitude of 7.67 \citep{bouchy05}.  At infrared wavelengths it is the brightest known star with a transiting planet, and its favorable planet/star radius ratio makes it ideal for a variety of detailed measurements \citep[see, for example,][]{dem06,grill07,knut07,tinn07,pont08,char08,swain08}.  On UT 2006 Oct. 28/29 we observed this planet continuously in the \emph{Spitzer} IRAC 8~\micron~bandpass for 33 hours, spanning slightly more than half of its orbit.  The high cadence of these data made it possible for us to fit the resulting phase curve with an ``orange-slice'' model for the planet consisting of twelve longitudinal strips of constant brightness.  The small size of the observed phase variation argued for highly efficient thermal homogenization between the two hemispheres, in contrast to the large day-night temperature differences inferred for $\upsilon$~Andromedae b and HD 179949b.

Although this longitudinal temperature map provides a wealth of information about the circulation within HD~189733b's atmosphere, its interpretation is complicated by the fact that the altitude of the atmospheric layer corresponding to the derived map depends on the atmospheric opacity at 8~\micron.  For wavelengths where the opacity is low the effective photosphere of the planet is located deep in the atmosphere, where the pressures and temperatures are correspondingly higher.  In their dynamical models \citet{coop05}, \citet{dobb08}, and \citet{show08} show that temperatures likely become increasingly homogenized at these higher pressures, as the radiative time scale increases relative to the advective time scale.  Thus we would expect that observations of the same planet at different wavelengths might show varying brightness contrasts between the day and night sides, depending on how deep into the atmosphere we are looking at each wavelength.  This also means that there is an inherent difficulty in comparing observations of one planet ($\upsilon$~Andromedae b) at 24~\micron~with observations of other planets (HD 179949b and HD 189733b) at 8~\micron.

In this paper we present new observations of the phase variation of HD~189733b at 24~\micron, spanning the same part of the planet's orbit as the previous observations at 8~\micron~described in Paper I.  These observations allow us to compare directly the properties of this planet's atmosphere at different wavelengths, and to search for wavelength-dependent differences that might indicate the relative opacities and corresponding depth of the planet's photosphere at each wavelength.  Models by \citet{coop05}, for example, predict that the light curves for hot Jupiters should vary in specific ways with increasing depth, with the day-night brightness contrast decreasing and the hot region on the day side advected increasingly far to the east.  Because these data were taken a year after our 8~\micron~observations, it is possible that dynamic weather patterns may have substantially altered the shape of the planet's phase curve \citep{cho03,cho08,rau08}.  If the observed light curve at 24~\micron~shows substantially different features, it would provide evidence for changing circulation patterns in the planet's atmosphere.  These observations also allow for a direct comparison between $\upsilon$~Andromedae b and HD~189733b, providing a definitive answer to the question of whether the inferred day-night temperature differences for these two planets indicate fundamental differences in their atmospheres or are simply the result of the differing opacities in these two bandpasses.

\section{Observations and Analysis}\label{obs}

We obtained 10,104 images of HD~189733 on UT 2007 October 25/26 using the \emph{Spitzer} MIPS 24~\micron~array \citep{wern04,rieke04} with a 10~s integration time.  Our observations spanned 35.5 hours, beginning 4.3 hours before the start of the transit and ending 2.8 hours after the end of the secondary eclipse.  There were two interruptions for data downloads, occurring approximately 1/3 and 2/3 of the way through the observations.  

\begin{figure}
\epsscale{1.2}
\plotone{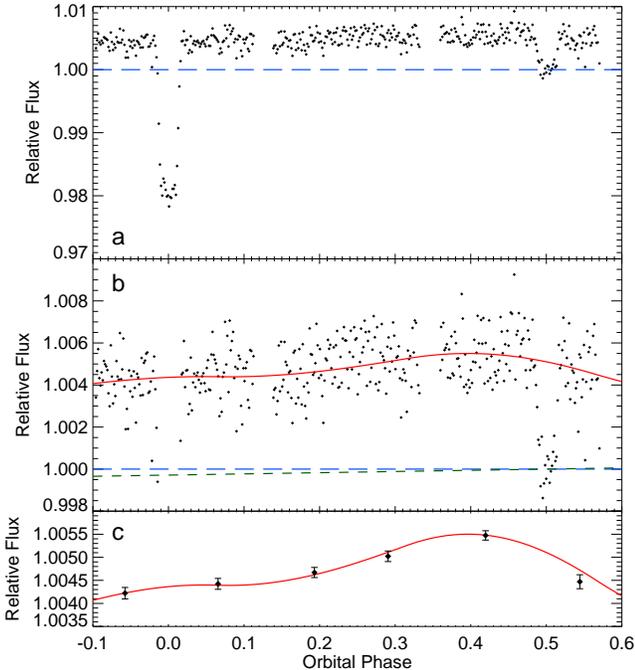}
\caption{Phase variation observed for HD~189733b by \emph{Spitzer} in the MIPS 24~\micron~bandpass, with transit and secondary eclipse visible.   The data are binned in 6.3 minute intervals.  In this Figure and Figure \ref{8micron_timeseries}, the stellar flux as measured at the center of the secondary eclipse has been normalized to unity (long-dashed line).  Panels {\bf a}, {\bf b} and {\bf c} show the same data, but in {\bf b} and {\bf c} the y axis is expanded to show the scale of the observed variation.  The bin size is also increased singificantly in {\bf c} to allow for a better comparison with the model fit.  The solid line in {\bf b} and {\bf c} is the phase curve for the best-fit four slice model in Figure \ref{map}, and the short-dashed line in {\bf b} shows the expected change in the star's flux as a result of the rotational modulation in the visibility of star spots over the period of the observations.  The flux from the planet after accounting for these spots would be the difference between the solid and short-dashed lines. 
\label{binned_timeseries}}
\end{figure}

The standard MIPS observing sequence dithers the target through 14 positions on the detector array (see \S8.2.1.2.1 of the \emph{Spitzer Observer's Manual}), cycling several times through seven vertically-offset scan-mirror (chop) positions on the left side of the array with a final observation at the starting position and then a nod that places the star on the right-hand side of the array before repeating the same sequence.  There are small differences in the apparent sensitivity at each position, and as a result we elect to treat each position as an independent data set in our analysis.  We discard a single position entirely, corresponding to the uppermost left position on the array.  This position falls within a few pixels of the position of a bright star, HD 350998, observed during the second nod position when HD 189733 is located on the right-hand side of the array.  HD 350998 is almost ten times brighter than our target in this bandpass, and as a result it is saturated in these images.  This saturation may produce undesirable effects that carry over into other pointing positions, and indeed we find that the time series for HD~189733b corresponding to this position differs noticeably from the other positions.  

In order to calculate the flux in each image, we first estimate the sky background from a \mbox{41 $\times$ 41} pixel box centered on the position of the star, excluding the pixels in a central \mbox{13 $\times$ 13} pixel region that includes both the star and its M dwarf companion\citep{bak06a}.  We iteratively trim all pixels more than $3\sigma$ away from the median value for this subarray, make a histogram of the remaining pixel values, and fit a Gaussian function to the central region of this histogram.

We note that the measured background drops by $1.5\%$ immediately after the telescope nod moving the star to a new set of positions (the pattern steps through positions $1-7$ for $5-6$ cycles, then nods and does the same for positions $8-14$).  \citet{dem05} noticed the same feature in their 24~\micron~observations of HD~209458.  In their data the measured flux from the star decreased along with the background flux, and they chose to take the ratio of the stellar to the background flux in order to remove this periodic drop.  We find no evidence of a corresponding decrease in the measured flux from HD 189733 in the images with $1.5\%$ lower background fluxes, therefore there is no need to apply the correction used by \citet{dem05}.  In either case, if we exclude these images from our final binned time series we obtain indistinguishable results.

\begin{figure}
\epsscale{1.2}
\plotone{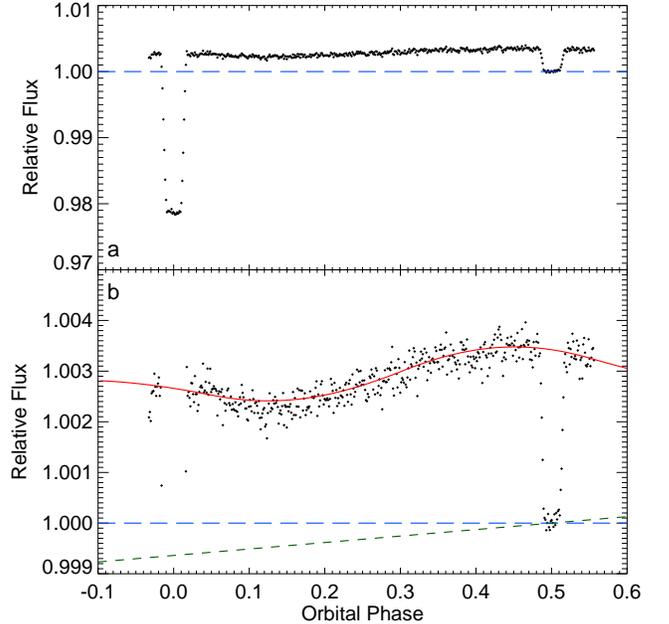}
\caption{Phase variation observed for HD~189733b by \emph{Spitzer} in the IRAC 8~\micron~bandpass, with transit and secondary eclipse visible.  The data are binned in 3.8 minute intervals.  These data were originally published in \citet{knut07}.  The stellar flux as measured at the center of the secondary eclipse has been normalized to unity (long-dashed line).  See Figure \ref{binned_timeseries} for a full description of the plotted quantities.
\label{8micron_timeseries}}
\end{figure}

We also see no evidence for the presence of the ``detector ramp'' effect \citep{char05,dem06,knut07,har07,knut08,char08} in these data.  This detector effect, which has been well-characterized in the IRAC 8~\micron~channel, causes the effective gain (and thus the measured flux) in individual pixels to increase over time.  The shape of the ramp depends on the illumination level of the individual pixel, with the most strongly illuminated pixels ($>250$ MJy Sr$^{-1}$) converging asymptotically to a constant value within the first two hours of observations, and the measured flux in the lowest-illumination pixels increasing linearly over time.  This effect is particularly  problematic for observations of phase variations, as the presence of this ramp can mimic a real rise in flux due to the planet's phase curve.  In Paper I we found that the correction for this effect increased the uncertainty in our estimate of HD~189733b's 8~\micron~night-side flux, which is based on data near the beginning of the observations, by a factor of five.  The IRAC 8~\micron~and MIPS~24~\micron~arrays are both Si:As detectors, and thus might be expected to behave similarly.  The IRS 16~\micron~array, which is also a Si:As detector, certainly does have a detector ramp \citep{dem06}.  It is not clear why the 24~\micron~data lacks this ramp, but we can immediately eliminate the shifting position of the star on the array as the explanation.  Although IRAC observations of eclipses typically use a single pointing \citep{char05,knut07,knut08,char08}, \citet{dem06} nodded the position of the star in their IRS 16~\micron~observations of HD~189733, which still shows a strong ramp.  Furthermore, both IRAC 8~\micron~and IRS 16~\micron~images also exhibit an even larger detector ramp in the measured values for the background fluxes over the period of the observations, independent of whether those images are nodded or not \citep{dem06,knut07,knut08,char08}.  We see no evidence for a ramp in the sky background in our 24~\micron~MIPS images, indicating that this effect is in fact absent from these images.  This is consistent with the conclusions of \citet{dem05}, who found no evidence for a ramp in their 6-hour MIPS observations.  It has been suggested that the higher background flux in the MIPS array may be saturating out this effect, but a comparison of MIPS 24~\micron~and IRS 16~\micron~peak up images indicates that the background flux in the 24~\micron~array in electrons per second is approximately twice that in the 16~\micron~array, which is not enough to explain the distinct behaviors of these two arrays.  The MIPS array is run at a higher bias voltage than IRS; this voltage is related to the detective quantum efficiency of the array, but as we do not fully understand the origin of the ramp it is difficult to say whether this might explain the differing behaviors of the MIPS and IRS photometry.

\begin{figure}
\epsscale{1.0}
\plotone{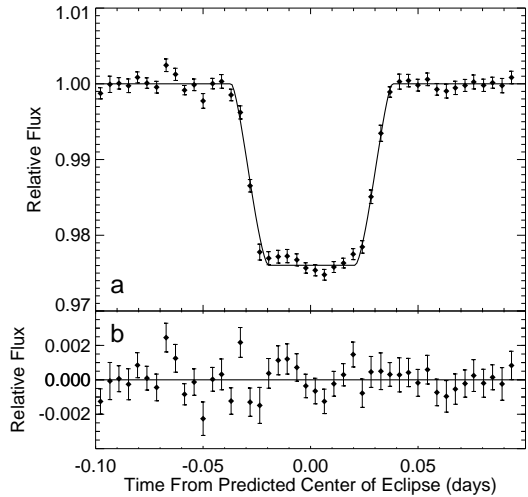}
\caption{This plot shows the transit {\bf a} of HD~189733b observed by \emph{Spitzer} in the MIPS 24~\micron~bandpass with best-fit curve overplotted including timing offsets.  Residuals from this fit are plotted in {\bf b}.  Data are binned in 6.3 minute intervals with error bars determined by the RMS variation in each bin divided by the square root of the number of points in the bin, and the out-of-transit points are normalized to one.
\label{transit_plot}}
\end{figure}

We estimate the flux from the star in each MIPS image as follows: first we subtract the background, then we fit the remaining flux in a circular region centered on the position of the star with a model point spread function.  We use a circular region (rounded to the nearest integer pixel) with a radius of six pixels and a MIPS model point spread function for a 5000 K point source\footnote{Available at http://ssc.spitzer.caltech.edu/mips/psf.html} for these fits.  This six-pixel radius is large enough to encompass the first Airy ring, and we find that increasing or decreasing the radius by one pixel does not affect the final time series.  We fix the position of this region for all of our fits at a given pointing position (meaning the boundaries of the region used for our fits do not shift with the position of the star, which varies by less than half a pixel at each pointing position over the period of our observations).   

To fit the observed point spread function, we interpolate our model to 100 times the resolution of the MIPS array and then rebin with the psf centered at the desired position, which we allow to vary in our fits.  This allows us to fit for the $x$ and $y$ position of the star to a resolution of 1/100th of a pixel.  The scatter on the final fitted positions is typically $\pm0.05$~pixels, five times larger than the resolution in our fits, so this is a reasonable choice.  We also fit for a constant scaling factor corresponding to the total flux, and use the error arrays generated by the standard \emph{Spitzer} pipeline to determine the relative weighting for individual pixels.  We note that the M-dwarf companion to HD~189733 is included within our subarray; we give zero statistical weight to the values within a 3 $\times$ 3 pixel box centered on the position of the companion in our fits.  This companion is located at a distance of 11\arcsec~from HD~189733 \citep{bak06a}, which places it on the outer edge of the first Airy ring for our target star.  Its flux is only 1/30th that of HD~189733 in this bandpass, thus a 3-pixel box is more than sufficient to eliminate any contribution from the companion. 

We flag bad pixels marked by the \emph{Spitzer} pipeline in our subarray and give them zero weight in our fits.  To find transient hot pixels, we collect the entire set of $702-840$ subarray images at a given pointing position, and calculate the median value and standard deviation at each individual pixel position.  We then step through the subarrays and mark outliers more than $3\sigma$ away from the median value for that pixel position as bad pixels in that image.  We find that $81\%$ of our images have one or fewer bad pixels, and $99\%$ have less than five bad pixels in the aperture used for our fits, which contains 113 pixels in total. This process reduces the number of large outliers in the final time series, although it does not eliminate such outliers completely.  We found that increasing our threshold for bad pixels to $4\sigma$ and then $10\sigma$ outliers produced comparable results with an increasing number of large outliers in the final time series.

\begin{figure}
\epsscale{1.0}
\plotone{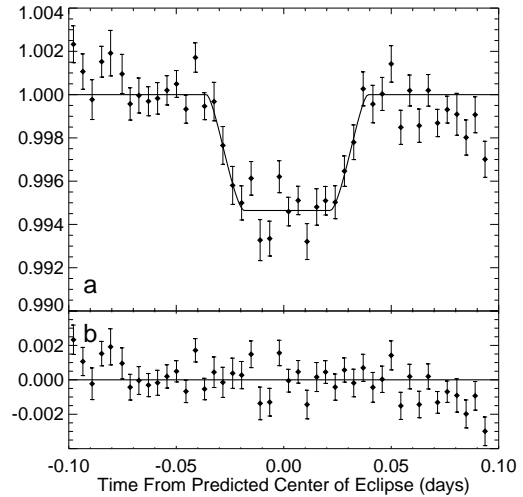}
\caption{This plot shows the secondary eclipse {\bf a} of HD~189733b observed by \emph{Spitzer} in the MIPS 24~\micron~bandpass with best-fit curve overplotted including timing offsets.  Residuals from this fit are plotted in {\bf b}.  Data are binned in 6.3 minute intervals with error bars determined by the RMS variation in each bin divided by the square root of the number of points in the bin, and the out-of-eclipse points are normalized to one.
\label{sec_eclipse_plot}}
\end{figure}

\begin{deluxetable*}{lrrrrcrrrrr}
\tabletypesize{\scriptsize}
\tablecaption{Best-Fit Eclipse Depths and Times \label{best_fit_param}}
\tablewidth{0pt}
\tablehead{
\colhead{Eclipse} & \colhead{Depth}  & \colhead{R$_{Planet}/$R$_{Star}$} & \colhead{Center of Transit (HJD)} & \colhead{O$-$C (s)\tablenotemark{b}} }
\startdata
8.0 \micron~Transit\tablenotemark{a} & $2.387\pm0.006\%$ & $0.1545\pm0.0002$ & $2454037.61196\pm0.00007$ & $-9\pm6~(\pm14)$\phantom{0...}\\
24 \micron~Transit & $2.396\pm0.027\%$ & $0.1548\pm0.0009$ & $2454399.24000\pm0.00019$ & $4\pm16~(\pm11)$\phantom{...}\\
8.0 \micron~Secondary Eclipse\tablenotemark{a} & $0.338\pm0.006\%$ & ~~ & $2454038.72294\pm0.00027$ & $116\pm23~(\pm6)$\tablenotemark{c,d}\\
24 \micron~Secondary Eclipse & $0.536\pm0.027\%$ & ~~ & $2454400.35033\pm0.00093$ & $65\pm80~(\pm11)$\tablenotemark{c}\phantom{.}\\
\enddata
\tablenotetext{a}{\citet{knut07}}
\tablenotetext{b}{This column gives the observed transit time minus the transit time calculated using the ephemeris derived in \S\ref{eclipse_fits} from a fit to 27 published transits of HD~189733b and the 24~\micron~transit from this paper. The uncertainties are set to the uncertainty in the observed transit time, while the values in parenthesis give the uncertainty in the predicted time.  The total uncertainty in the O$-$C values is the sum of these two values.}
\tablenotetext{c}{Predicted secondary eclipse times are defined as $T_c+0.5P+30$~s, where the additional 30 s delay accounts for the light travel time in the HD~189733 system \citep{loeb05}.}
\tablenotetext{d}{The predicted time for this eclipse is calculated using the $T_c$ determined from the 8~\micron~transit instead of the value quoted in \S\ref{eclipse_fits}, as this produces a more accurate prediction.}
\end{deluxetable*}

After producing a time series for each pointing position, we iteratively select and trim outliers greater than $3\sigma$ to remove any remaining points affected by transient hot pixels.  We include data spanning the transit and secondary eclipse in this iterative trimming process, but we first divide the time series by the best-fit transit and secondary eclipse light curves as determined in \S\ref{eclipse_fits} before selecting $3\sigma$ outliers.  After trimming these outliers from the original time series (including eclipses) we divide the trimmed time series by its median value and combine all of the pointing positions into a single time series consisting of 9,243 points, 91\% of the original total.  See Figure \ref{binned_timeseries} for the resulting 24~\micron~light curve and Figure \ref{8micron_timeseries} for the comparable 8~\micron~light curve from Paper I.  We set the uncertainties for each individual point equal to the standard deviation of this combined time series after the end of the secondary eclipse. 

\subsection{Fitting the Eclipses}\label{eclipse_fits}

We fit the transit and secondary eclipse using a  Markov Chain Monte Carlo (MCMC) method \citep[see, for example][]{ford05,winn07b} with $10^6$ steps.  We initialize the chain with the best-fit parameters determined from a $\chi^2$ minimization routine and add small random perturbations to these values to ensure that the chain explores the correct region of parameter space.  We calculate our transit and secondary eclipse light curves using the equations from \citet{mand02} for the case with no limb-darkening.  \citet{bea08} found that accounting for the effects of limb-darkening increased the resulting best-fit transit depth for HD~189733b by $0.02\%$ in the 5.8~\micron~IRAC bandpasses; we would expect that the effects of limb-darkening would be much smaller in the MIPS 24~\micron~bandpass.  As a test, we repeat our transit fit with a single linear limb-darkening coefficient as an additional free parameter, and find that the best-fit transit depth increases by $0.017\%$ or 0.6 $\sigma$.  Our best-fit value for the limb-darkening coefficient is $0.022\pm0.077$, indicating that our data are entirely consistent with a limb-darkening coefficient of zero.  As a result, we elect to fix the limb-darkening to zero in all of our subsequent fits.

Our free parameters in the fit include a constant scaling factor, the transit time, and the transit depth.  After running the chain, we search for the point where the $\chi^2$~value first falls below the median of all the $\chi^2$~values in the chain (i.e. where the code had first found an excellent fit), and discard all steps up to that point.  We take the median of the remaining distribution as our best-fit parameter, with errors calculated as the symmetric range about the median containing 68\%~of the points in the distribution.  The distribution of values was very close to symmetric in all cases, and there did not appear to be any strong correlations between variables.  Figures \ref{transit_plot} and \ref{sec_eclipse_plot} show the binned data with best-fit transit and secondary eclipse curves overplotted.  Best-fit eclipse depths and times are given in Table \ref{best_fit_param}.  As a check, we also repeated the same fits using a standard downhill simplex $\chi^2$ minimization routine, and obtained equivalent best-fit parameters.

Because the 24~\micron~transit is not as well-constrained as the 8~\micron~observations described in Paper I, we set the inclination to its best-fit value from the fit to the 8~\micron~transit.  We then fit for the depth of the transit at 24~\micron, which is proportional to the square of the ratio of the planetary and stellar radii, and the transit time.  We find that this 24~\micron~transit depth differs from the 8~\micron~transit depth from Paper I by $0.3\sigma$.  We perform the same fit for the secondary eclipse, allowing both the depth and timing to vary independently and set other parameters to their best-fit values from Paper I.  We find a relative depth of $0.536\pm0.027\%$.  This is consistent with the previous value of $0.598\pm0.038\%$ from a 24~\micron~eclipse observed in 2005 \citep{char08}, at a level of $1.3\sigma$.  

There have been several new high-precision observations of transits of HD~189733b published in the past year, including HST ACS observations \citep{pont07} and the \emph{Spitzer} 8~\micron~transit from Paper I.  Rather than using the previously published ephemeris from \citet{winn07a}, which does not include these recent observations, we derive a new ephemeris from a fit to all 27 previously published transits \citep{bak06b,winn07a,knut07,pont07}.  This new ephemeris has a central transit time $T_c=2454399.23990\pm0.00017$~HJD and a period $P=2.21857578\pm0.00000080$ days.  Using this new ephemeris we find that the 24~\micron~transit occurs $9\pm22$~s later than predicted.  Next we repeat this fit including the 24~\micron~transit, and find a central transit time $T_c=2454399.23995\pm0.00013$~HJD and a period $P=2.21857597\pm0.00000060$ days ($\pm52$ ms).  Using this ephemeris, we find that the secondary eclipse occurs $64\pm81$~s later than the predicted time, which is defined as $T_c+0.5P+30$~s \citep[the additional 30~s delay accounts for the light travel time between the planet and star, as calculated from][]{loeb05}.  Our timing precision is not sufficient to confirm or reject the $120\pm24$~s delay in the time of the secondary eclipse reported in Paper I.

\begin{deluxetable}{lrrrrcrrrrr}
\tabletypesize{\scriptsize}
\tablecaption{Comparison of the Minimum and Maximum Planet-Star Flux Ratios \label{flux_table}}
\tablewidth{0pt}
\tablehead{
\colhead{Parameter} & \colhead{8~\micron}  & \colhead{24~\micron} }
\startdata
$F_{min}$ & $0.219\pm0.024\%$ & $0.416\pm0.027\%$ \\
$F_{max}$ & $0.342\pm0.006\%$ & $0.550\pm0.0027\%$ \\
$F_{min}/F_{max}$ & $64\pm7\%$\phantom{000.}& $76\pm3\%$\phantom{000.}\\
$F_{min,corr}$\tablenotemark{a} & $0.261\pm0.025\%$ & $0.443\pm0.027\%$ \\
$F_{min,corr}/F_{max}$\tablenotemark{a} & $76\pm7\%$\phantom{000.} & $81\pm3\%$\phantom{000.} \\
\enddata
\tablenotetext{a}{This gives the minimum planet/star flux ratio after subtracting the estimated contributions from the star spots.  The maximum planet/star flux ratios are not affected by these spots.}
\end{deluxetable}

\subsection{Fitting the Phase Curve}\label{phase_fits}

We fit the observed phase variation with an ``orange-slice'' model of the planet consisting of $N_{\rm slices}$ longitudinal slices with a uniform intensity in both longitude and latitude \citep{cow08}.  We find that smoothing this step function does not significantly change the resulting light curve, provided the total flux from each slice and its brightness-weighted longitude are unchanged. The slices are centered at ${\phi_{0}, \phi_{0}+\Delta \phi, \phi_{0}+ 2\Delta \phi, \dots}$, where $0 \le \phi_{0} < \Delta \phi$ is a free parameter in the fit. The model therefore has $N_{\rm slices}+1$ free parameters. The phase offset is necessary because it is possible to determine $\phi_{0}$ to better than $\pm \Delta \phi/2$ since the projected area of a slice peaks when the center of the slice is facing the Earth.

We test a series of models with either 2, 3, 4, or 6 slices. The best-fit model parameters and the associated 1$\sigma$ uncertainties are determined using a MCMC method as described in \S\ref{eclipse_fits}.  We initialize this chain using the best-fit values from the Levenberg-Marquardt fit.  The reduced $\chi^{2}$ is fairly insensitive to changes in the number of slices, but the uncertainty in the intensity of each individual slice increases with $N_{\rm slices}$. We elect to use the four-slice model fit for our final analysis, as this represents the best compromise between the degree of spatial resolution and the uncertainties in the flux from each individual slice. 

For all of these models, we find that the brightest region on the planet is located to the east of the substellar point, consistent with a peak in the integrated phase function occurring before the start of the secondary eclipse.  For our two-slice model the brightest slice is centered $31\pm9$\degr~to the east of the substellar point, and in the four-slice model it is centered $22\pm8$\degr~to the east.  The peak in the integrated light curve for the four-slice model, corresponding to the maximum hemisphere-averaged flux from the planet, occurs at an orbital phase of $0.396\pm0.022$ or a central meridian longitude $37\pm8$\degr~east of the substellar point (note that the longitude of the brightest slice and the longitude of the brightest hemisphere are not the same for asymmetric longitudinal flux distributions).  With a difference of 2.2$\sigma$, this is mildly inconsistent with the 8~\micron~phase curve, which reaches its maximum value at an orbital phase of $0.456\pm0.017$ or a central meridian longitude $16\pm6$\degr~east of the substellar point. 

 \begin{deluxetable}{lrrrrcrrrrr}
\tabletypesize{\scriptsize}
\tablecaption{Comparison of the Minimum and Maximum Hemisphere-Averaged Brightness Temperatures \label{bright_temp}}
\tablewidth{0pt}
\tablehead{
\colhead{Parameter} & \colhead{8~\micron \tablenotemark{a}}  & \colhead{24~\micron} }
\startdata
$T_{max}$ & $1258\pm11$~K & $1220\pm47$~K \\
$T_{min}$ & $1011\pm51$~K & $984\pm48$~K \\
$T_{max}-T_{min}$ & $247\pm51$~K & $236\pm48$~K \\
$T_{min,corr}$\tablenotemark{b} & $1098\pm51$~K & $1032\pm48$~K \\
$T_{max}-T_{min,corr}$\tablenotemark{b} & $160\pm51$~K & $188\pm48$~K \\
\enddata
\tablenotetext{a}{These temperatures are higher than the values published in Paper I because we use a Kurucz atmosphere model specific to HD~189733 (available at http://www.kurucz.cfa.harvard.edu/stars) to determine the flux from the star rather than interpolating from a grid of atmosphere models.}
\tablenotetext{b}{This gives the minimum hemisphere-averaged brightness temperatures after subtracting the estimated contributions from the star spots (the maximum temperature estimates are not affected by these spots).}
\end{deluxetable}

Because the region of the planet located 90\degr~east of the antistellar point was visible only briefly at the beginning and end of the observations, it is not as well constrained by these data as the other regions of the planet.  For the 24~\micron~data, our best-fit models indicate that the coldest region on the planet may be located anywhere from 30\degr~east (from a two-slice fit) to 90\degr~east (from a four-slice fit) of the antistellar point.   This would correspond to a minimum in the planet's phase curve that occurred before the start of our observations, which explains why its location is poorly constrained by these fits.  There is a local minimum in the phase curve for the best-fit four-slice model occurring at an orbital phase of 0.1 or a central meridian longitude $40$\degr~west of the antistellar point, but this feature is not statistically significant and we are unable to confirm or exclude the existence of a minimum comparable to the one observed in the 8~\micron~data.   

Based on the four-slice fit, we conclude that the observed phase curve has a maximum in flux of $1.00550\pm0.00010$ where the stellar flux as measured at the center of the secondary eclipse has been normalized to unity.  The minimum flux in this fit is $1.00349\pm0.00038$, but this involves an extrapolation to a time prior to the start of our observations.  We use a simpler two-slice model to estimate the minimum flux and find a value of $1.00416\pm0.00011$, which is consistent with the median flux value prior to the start of the transit.  When we estimate the brightness temperatures we must include the uncertainty in the depth of the secondary eclipse, as this measurement determines the baseline flux contribution from the star.  This increases the uncertainties in both the minimum and maximum flux estimates to $\pm0.0027$.  Taking the difference of the maximum four-slice and minimum two-slice fluxes, we find an increase of $0.133\pm0.015\%$ in the measured flux from this system.  This corresponds to a night-side flux from the planet that is $76\pm3\%$ of the day-side flux, where the uncertainty includes the propagated uncertainties in the minimum and maximum fluxes as well as the depth of the secondary eclipse, which is used to determine the planet-star flux ratio.  This is consistent at a level of 1.6$\sigma$~with the results of our previous 8~\micron~observations, which found that the night-side flux was $64\pm7\%$ of the day-side flux.  The uncertainties are comparable for the 24~\micron~light curve despite the increased scatter in the data because the correction for the detector ramp in the 8~\micron~bandpass adds significant uncertainty to the 8~\micron~night-side flux (see \S\ref{obs} and Paper I for more detailed explanations of the 8~\micron~ramp correction).  The 24~\micron~night-side flux matches the prediction by \citet{bar08}, who used the planet's day-side broadband emission spectrum \citep{char08} and 8~\micron~phase curve to predict a night/day flux ratio of $75\%$ at 24~\micron.

\begin{figure}
\epsscale{1.2}
\plotone{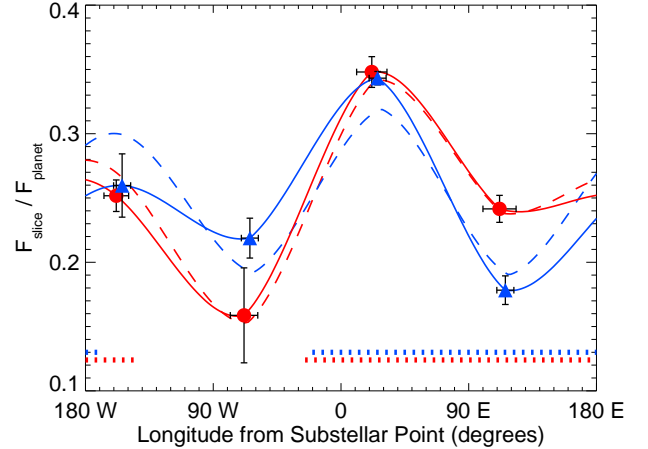}
\caption{Brightness estimates for four longitudinal strips on the surface of the planet at 8~\micron~(blue triangles) and 24~\micron~(red circles), where the brightness values are given as a percentage of the total flux from the planet.  The solid overplotted lines are spline interpolations of these four-slice fits, and the dashed lines show the same model fits after correcting for the effects of star spots.  The horizontal dotted lines at the bottom of the plot indicate the range of central meridian longitudes (longitudes viewed face-on) visible during our IRAC (upper line) and MIPS (lower line) observations.
\label{map}}
\end{figure}

In order to facilitate comparisons between the model fits at 8~and 24~\micron, we repeat our four-slice fit using the 8~\micron~phase curve from Paper I.  The resulting four-slice fit is plotted together with the 24~\micron~model in Figure \ref{map}, and the corresponding integrated phase curve is shown in Figure \ref{8micron_timeseries}.  We also calculate the minimum and maximum hemisphere-averaged brightness temperatures (see Table \ref{bright_temp}) corresponding to the minimum and maximum fluxes in the 8 and 24~\micron~light curves from Table \ref{flux_table}.  To determine these brightness temperatures, we use a Kurucz atmosphere model for the star\footnote{Available at http://kurucz.harvard.edu/stars.html} \citep{kurucz79,kurucz94,kurucz05} and assume a Planck function for the planet.  We take the ratio of these two functions and integrate over the IRAC 8~\micron~and MIPS 24~\micron~bandpasses to determine the planet-star flux ratio in each bandpass, and solve for the temperature that matches the flux ratio observed in each bandpass.  Note that our estimate of the planet's brightness temperature is dependent on our choice of the atmosphere model for the star; in Paper I we interpolated from a grid of models, whereas in this paper we use a model specific to HD 189733.  This new model has a slightly higher effective temperature and correspondingly higher 8~\micron~flux, and as a result our estimates of the brightness temperatures for the planet in the 8~\micron~bandpass are higher than the values given in Paper I.

\subsection{Effects of Starspots}

HD 189733 is an active K0 star \citep{bouchy05}, which has been observed to vary by $\pm1.5$\%~at visible wavelengths \citep{winn07a,henry08,croll07}.  Because the spots on this star have an effective temperature approximately 1000~K cooler than that of the stellar photosphere \citep{pont08}, the amplitude of these variations scales approximately as the ratio of two blackbodies, which would imply that variability from spots should have a much smaller amplitude at 24~\micron.  However, even a small variation in the star's flux during the period of our observations might contribute significantly to the observed signal.  In order to characterize the behavior of these spots, we obtained simultaneous ground-based observations of HD~189733 using the 1.2 m telescope at the Fred Lawrence Whipple Observatory (FLWO) and one of the Tennessee State University 0.8~m automated photometric telescopes (APT) at Fairborn Observatory.  Although the FLWO observations span only the week surrounding our MIPS observations, the APT observations are part of a long-term monitoring program \citep{henry08} and extend over more than a year, including the times of both our 8~and 24~\micron~\emph{Spitzer} data.  These observations allow us to determine the behavior of the star during these two periods.

\subsubsection{FLWO Photometry}

We used KeplerCam on the 1.2 m telescope at the Fred L. Whipple Observatory on Mt. Hopkins, Arizona to obtain Sloan \emph{g} and \emph{z} photometry of HD 189733 on seven consecutive nights beginning on UT 2007 October 22.  This instrument has a 23\arcmin.1$\times$23\arcmin.1 field of view, allowing us to obtain photometry of a number of bright comparison stars simultaneously with our target \citep[see][for more detailed information on observations of HD~189733 using this instrument]{winn07a}.  HD~189733 was setting early in the evening, so these observations were taken during a brief period at the beginning of each night.   We used a single master flat constructed from the dome flats taken over all nights to correct the images.  Images were intentionally defocused to avoid saturating the array during our 5~s exposures, and we discarded any images where the point spread function for HD~189733 contained saturated pixels.  We used aperture photometry with IRAF's PHOT task and an aperture radius of 35 pixels or 23.5\arcsec~to estimate the flux from HD~189733 in each image.  Increasing the size of this aperture produced a corresponding increase in the amount of noise from the sky background, while decreasing it led to larger systematic variations.  We estimated the level of sky background using an annulus with a radius $35<r<60$ pixels. 

\begin{figure}
\epsscale{1.2}
\plotone{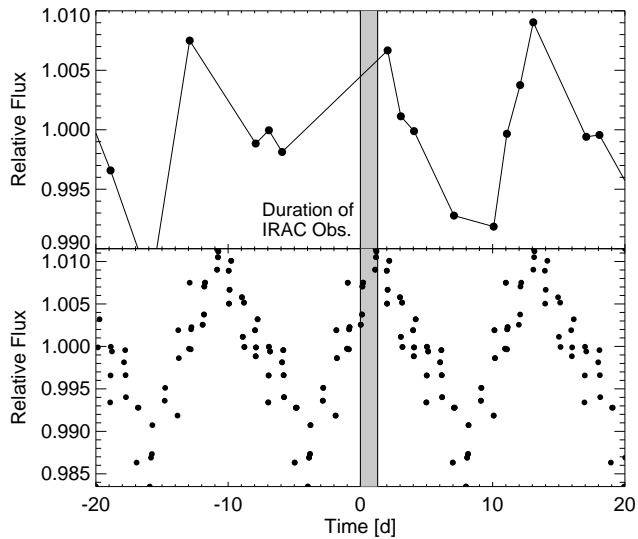}
\caption{Relative flux variation for HD~189733 observed with the T10 APT in the Str\o mgren $y$ band encompassing the time of our 8~\micron~IRAC observations (upper panel).  Vertical lines indicate the start and end times for our \emph{Spitzer} observations, which occurred during a period of increasing stellar flux.  Variations are caused by rotational modulation in the visibility of star spots with a rotation period of $11.953\pm0.009$ days \citep{henry08}.  The lower panel shows the variations observed during a several-month period around the 8~\micron~\emph{Spitzer} observations phased by the rotation period of the star.
\label{starspots}}
\end{figure}

We corrected for variations in atmospheric transmission and instrument efficiency using a set of comparison stars visible in the images, iteratively discarding comparison stars with light curves that appear to differ significantly from the average.  There were ten comparison stars in the final iteration, with a relative total flux three times greater than that of HD~189733.  We estimated the mean flux for HD~189733 each night relative to the ensemble average of these ten calibrators, and set the uncertainties equal to the RMS variation in these relative fluxes divided by the square root of the number of images.  The resulting flux values in the \emph{g} bandpass for each night are plotted in Figure \ref{starspots}.  We obtained similar measurements in \emph{z} but most of the images from the latter part of the week had saturated pixels, and the limited time coverage of the remaining images meant that they were not useful for our analysis.

\subsubsection{APT Photometry}

We obtained observations of HD~189733 in Str\"omgren \emph{b} and \emph{y} filters over a span of several months surrounding our 8 and 24~\micron~\emph{Spitzer} observations from an ongoing monitoring program carried out with the T10 0.8 m APT at Fairborn Observatory in southern Arizona \citep{henry99,eat03}.  In these observations the telescope nodded between HD 189733 and three comparison stars of comparable or greater brightness as described in \citet{henry08}.  Because we are ultimately interested in the behavior of the starspots at longer wavelengths, we elect to use the \emph{y} band photometry for our analysis (see Figure \ref{starspots}).  We note that \citet{henry08} published similar observations of HD 189733 spanning the period of our 8~\micron~observations of this system; this allows us to determine the behavior of the star during this period as well.

\subsubsection{Spitzer IRAC 8~\micron~Photometry}\label{spitzer_phot}

\begin{figure}
\epsscale{1.2}
\plotone{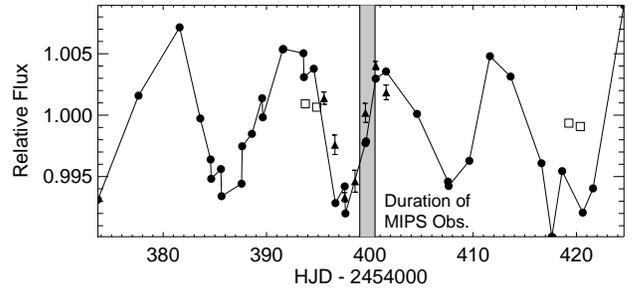}
\caption{Relative flux variation for HD~189733 observed with the T10 APT in the Str\o mgren $y$ band (filled circles), with the FLWO 1.2m in the Sloan $g$ band (filled triangles), and \emph{Spitzer} IRAC 8~\micron~band (open squares) around the time of our 24~\micron~\emph{Spitzer} MIPS observations.  Vertical lines indicate the start and end times for the MIPS observations, which occurred during a period of increasing stellar flux.  The first and fourth IRAC 8~\micron~points correspond to secondary eclipses, while the second and third points are transits.  The average over these four points has been set equal to one in this plot.  The first and second IRAC observations are located during a period of maximum stellar flux and the third and fourth IRAC observations are located during a period of minimum stellar flux.  The decrease in flux over the time of these observations provides a direct estimate of the size of the variations induced by spots in the 8~\micron~IRAC bandpass.
\label{spitzer_starspots}}
\end{figure}

The FLWO and APT photometry indicates that the star is increasing in flux during both of our observations, and we can use this information to estimate the amplitude of the corresponding variations in the 8 and 24~\micron~bandpasses.  We scale these variations to infrared wavelengths using unpublished \emph{Spitzer} observations of two transits and two secondary eclipses of HD~189733b in the IRAC 8~\micron~bandpass (program GO 40238, PI E. Agol), taken during the weeks immediately before and after our MIPS observations.  These observations include a secondary eclipse on UT 2007 October 20, a transit observed on UT 2007 October 21, a transit observed on UT 2007 November 14, and a secondary eclipse observed on UT 2007 November 15 (see Figure \ref{spitzer_starspots}).  The first transit/secondary eclipse pair occur during the stellar maximum immediately preceding our MIPS observations, and the second transit/secondary eclipse pair occur during the stellar minimum a little more than three weeks later.

We estimate the flux from the star in individual images by summing the flux within circular apertures with radii ranging from 3.4 to 7 pixels centered on the position of the star.  For apertures smaller than 3.4 pixels there are flux losses correlated with the position of the star on the array, and apertures larger than seven pixels become increasingly noisy.  We determine this position using a weighted sum of the fluxes over a 7$\times$7 pixel box centered on the approximate position of the star, following the methods described in \citet{knut07,knut08} and \citet{char08}.  We subtract a background from each image determined by fitting a Gaussian function to a histogram of pixels in the corners of the subarray images where flux from the star as minimal.  We do not apply an aperture correction, as we are only interested in estimating relative changes in the flux from the star over time.

We determine the decrease in the flux from the star during these two epochs by comparing data from the two transits and the two secondary eclipses separately.  This is because we expect the flux from the planet to vary as it moves through its orbit, but this effect cancels out if we compare data from the same region of the phase curve (i.e. immediately after the transit or immediately after the secondary eclipse).  This assumes that the planet does not experience significant weather-related variability, such as that suggested by \citet{cho03,cho08} and \citet{rau08}.  The overall consistency in the observed features of our 8 and 24~\micron~light curves, which are separated in time by approximately one year, indicates that weather-related variability in the observed fluxes is probably minimal.  Our 24~\micron~secondary eclipse depth is also consistent with a 24~\micron~eclipse observed in 2005 \citep[see \S\ref{eclipse_fits} and][]{char08}, placing an additional constraint on the variability.

We find that the median flux measured after the end of the transit decreased by $0.15\%$ between the two epochs and the median flux after the end of the secondary eclipse decreased by $0.20\%$ between the two epochs (see Figure \ref{spitzer_starspots}), where each of these measurements was averaged over apertures ranging from 3.4 to 7 pixels in radius.  We use a range of aperture sizes in order to test whether the detector ramp described in \S\ref{obs}, which causes the effective gain of the detector to increase as a function of time, is affecting our result here.  Pixels that are more strongly illuminated have both a shorter characteristic time scale and a smaller relative amplitude for the ramp (on the order of $1\%$ for high-illumination pixels, versus $10\%$ for the lowest-illumination pixels).  Thus, over the 5-6 hour time frame of these eclipse observations, the high-illumination pixels at the center of the star's point spread function have already converged to a constant value, whereas the lower-illumination pixels near the edge of the aperture are contributing the majority of the observed ramp.  We find that the median flux after the end of the transit decreases by $[0.17\%, 0.12\%, 0.18\%, 0.14\%]$ between the two epochs for apertures with radii of $[3.4, 3.5, 5.0, 7.0]$ pixels, respectively.  The consistency of these values over a range of aperture sizes indicates that the detector ramp described in \S\ref{obs} is effectively removed in this ratio.  We have no particular reason to prefer one aperture size over another, and there may be other effects at work.  Therefore, we elect to average over these four apertures.  We use the same method to estimate the decrease in the star's flux over the epoch bounded by the two secondary eclipse observations, and find values that are similarly consistent over a range of apertures.  

As a second test we take the ratio of the median fluxes over the entire observations, trimming only the first hour of data where the ramp is steepest.  This further tests whether the detector ramp is influencing our results, as we have now added in more data at earlier times when the detector ramp is larger.  We again find consistent results over a range of apertures, with the star's flux decreasing by $0.11\%$ on average between one transit and the next, and by $0.17\%$ on average between the two secondary eclipses.  Together with the two values derived solely from data after the ends of the eclipses, these values represent four independent estimates of the decrease in stellar flux over this period.  Combining these values, we find that the flux from the star decreases by $0.16\pm0.02\%$ in the IRAC 8~\micron~bandpass between these two epochs, while it decreases by 1.3\%~over the same period in the Str\"omgren \emph{y} bandpass. 

\subsubsection{Scaling the Starspots to \emph{Spitzer} Wavelengths}

As shown in Figure \ref{starspots}, we have APT observations in the Str\"omgren \emph{y} bandpass spanning the period of both our MIPS 24~\micron~phase variation observations on UT 2007 Oct. 25/26 and our previous IRAC 8~\micron~phase variation observations on UT 2006 Oct. 28/29.  The FLWO \emph{g} and APT \emph{y} band observations are in good agreement and show that the flux from the star is increasing during the times of our 8 and 24~\micron~\emph{Spitzer} observations (these are the grey shaded regions in Figure \ref{starspots}).  We determine the increase in the star's flux during the period of our IRAC 8~\micron~observations by fitting a linear function of time to the phased Str\"omgren \emph{y} data plotted in the bottom panel of Figure \ref{starspots}, beginning at the flux minimum before the start of our IRAC observations and ending at the flux maximum shortly after the end of these observations.  From this fit we find that the star's flux increased by $0.0196\pm0.0002\%$ per hour during our IRAC 8~\micron~observations in the Str\"omgren \emph{y} bandpass. Unlike the earlier data from 2006 spanning the IRAC 8~\micron~observations, the variations plotted in the top panel of Figure \ref{starspots} during the period of our MIPS observations in 2007 do not phase well, indicating that the properties of the spots are varying during this epoch.  We estimate the increase in the star's flux during the MIPS observations by fitting a linear function of time to the unphased Str\"omgren \emph{y} data, beginning at the flux minimum before the start of our MIPS observations and ending at the flux maximum shortly after the end of these observations. We find that the star's flux increased by $0.011\pm0.001\%$ per hour in the Str\"omgren \emph{y} bandpass during our MIPS observations.  

To determine the star's contribution to the flux variations observed in the 8 and 24~\micron~bandpasses, we must scale the observed changes in the Str\"omgren \emph{y} bandpass to reflect the decreased contrast of these starspots relative to the star's photosphere at infrared wavelengths.  For the IRAC observations, this scaling is simple: as discussed in \S\ref{spitzer_phot}, a decrease of 1.3\% in \emph{y} corresponded to a decrease of $0.16\pm0.02\%$ at 8~\micron.  Applying the same scaling to the observed increase in Str\"omgren \emph{y} during our 8~\micron~\emph{Spitzer} observations, we estimate the star increased in flux by $0.0024\pm0.0003\%$ per hour at 8~\micron.  The total increase in flux observed in this bandpass was $0.12\pm0.02\%$ over 17.6 hours.  This implies that the star contributed $0.042\pm0.005\%$ to the observed increase in flux, one-third of the total signal.  Higher cadence MOST data obtained several months prior to these observations \citep{croll07} indicates that the increase in flux during this part of the star's rotation is effectively linear; in either case the small size of the star's contribution makes it very unlikely that the minimum and maximum in the 8~\micron~light curve can be explained by the effects of star spots.

To estimate the contribution of the starspots to our MIPS 24~\micron~data, we first scale the star's $0.011\pm0.001\%$ per hour increase in Str\"omgren \emph{y} to the equivalent value of $0.0013\pm0.0002\%$ per hour in the \emph{Spitzer} 8~\micron~bandpass.  Next, we estimate how the effects of the spots scale between the 8~and 24~\micron~bandpasses.  The precise scaling depends on the temperatures of the spots relative to the effective temperature of the star.  Previous HST ACS observations of HD 189733 have established that these spots have temperatures between 4000-4500~K \citep{pont08}, and we use this temperature range in our analysis.  We estimate the relative decrease in flux $df\left(\lambda\right)$ from these spots as the difference between spectra from a grid of model atmospheres \citep{kurucz79,kurucz94,kurucz05} with a temperature of 5000 K and either 4500 or 4000 K:

\begin{equation}
df\left(\lambda\right)=\frac{f_\star\left(\lambda\right)-f_{spot}\left(\lambda\right)}{f_\star\left(\lambda\right)}
\end{equation}

The effective temperature of the star is $5050\pm50$~K \citep{bouchy05}, so a 5000~K model is a reasonable choice for the star.  We take the weighted average of $df\left(\lambda\right)$ over the IRAC 8~\micron~bandpass and then over the MIPS 24~\micron~bandpass, and find that a $0.0013\pm0.0002\%$ per hour increase in flux at 8~\micron~would correspond to an increase of $0.0011\pm0.0002\%$ per hour in the MIPS 24~\micron~bandpass.  The observed increase in flux is $0.133\pm0.015\%$ over 25 hours, so this implies that the star contributes $0.027\pm0.004\%$ or approximately one-fifth of this increase.  As a check we recalculate this scaling using Planck functions and find that we obtain indistinguishable results.

In both the 8 and 24~\micron~bandpasses, accounting for the effects of star spots results in a slightly warmer minimum hemisphere-averaged brightness temperature, but does not otherwise alter our conclusions.  This is because the maximum hemisphere-averaged brightness temperature is set by the depth of the secondary eclipse (which gives the total flux from the planet at that point relative to the flux from the star) relative to the maximum in the phase curve, and the interval between these two events is relatively short.  The minimum hemisphere-averaged brightness temperature, on the other hand, is set by the changes in the observed flux over a much longer period of time where the effects from star spots are increasingly important.  We give the minimum planet/star flux ratios for both bandpasses after correcting for the effects of these spots in Table \ref{flux_table}, and the corresponding minimum hemisphere-averaged brightness temperatures in Table \ref{bright_temp}.

\section{Discussion}
\subsection{Day and Night Atmospheric Structure}
The picture that emerges from our 24~\micron~observations broadly matches the situation we previously inferred from the planet's 8~\micron~light curve in Paper I.  In radiative equilibrium, tidally locked hot Jupiters should exhibit day-side temperatures at the photosphere that exceed 1300 K and night-side temperatures as low as 200--300 K, implying a day-night temperature difference exceeding 1000 K \citep[e.g.][]{show08, bar05}.  In contrast to this reference state, we find that the planet exhibits similar day- and night-side brightness temperatures at 24~\micron, with a night side only modestly colder than the day side.  In fact, the difference between the maximum and minimum hemisphere-averaged temperatures that we infer at 24~\micron, $236\pm48\,$K, is indistinguishable from our previously estimated 8~\micron~value of $247\pm51\,$K.  Accounting for the effects of star spots results in a slightly warmer minimum hemisphere-averaged temperature at both wavelengths, but does not otherwise affect our conclusions (see Table \ref{bright_temp}).

Relative to these values, current three-dimensional circulation models over-predict the day-night flux variations at both 8 and 24~\micron~\citep{fort06,show08}.  These same models indicate that the day-night temperature difference should increase with altitude \citep{coop05,show08,dobb08}, but this change is modest over the factor of $2-3$ variation in the pressures that are likely sensed by these two bandpasses.  The small size of the observed flux variations in both the 8~and~24~\micron~light curves indicates that the circulation efficiently transports thermal energy from the day side to the night side over the range of pressures spanned by the 8~\micron~and 24~\micron~photospheres, leading to moderately (though not completely) homogenized temperatures between the day side and night side.  

\begin{figure}
\epsscale{1.2}
\plotone{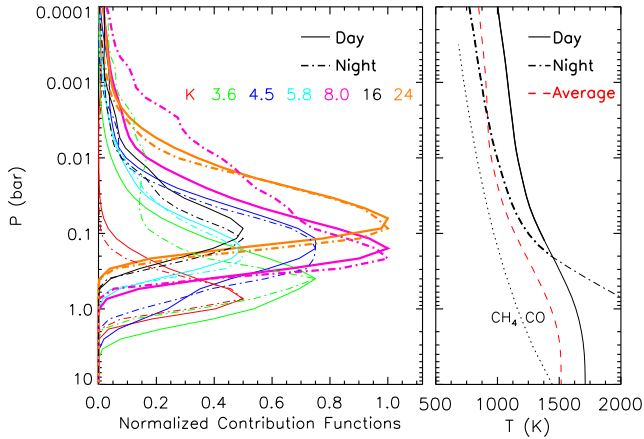}
\caption{\emph{Left panel:} Normalized contribution functions for 1D model atmospheres of HD~189733b.  The day side is plotted with solid curves, while the night side uses dash-dot curves.  Band-average contribution functions are shown for various \emph{Spitzer} bands and at $K$.  For clarity, the 8 and 24~\micron~curves are normalized to 1.0, while the 3.6 and 4.5~\micron~bandpasses are normalized to 0.75 and the remaining bandpasses are normalized to 0.5.  \emph{Right panel:}  Atmospheric pressure-temperature profiles for the day and night hemispheres.  For comparison, in dashed red is shown a 1D planet-wide average profile, which is computed like the day-side profile, but with 1/2 the stellar flux, due to the day/night average.  The night side profile is that of 1500 K isolated object, which has a much warmer interior than that of HD~189733b, and hence the atmosphere at high pressure becomes unrealistically hot.  The dotted curve shows where CH$_4$ and CO are equal in abundance.
\label{contrib}}
\end{figure}

There is a second possible explanation for the similarities between the 8 and 24~\micron~light curves.  If both bandpasses sense similar atmospheric pressures then we would expect to see the same features in both light curves, regardless of how the atmospheric circulation varied with pressure.  We investigate this issue by computing day-side average and non-irradiated night-side 1D models of HD 189733b that assume solar metallicity, negligible TiO/VO opacity due to condensation, and neglect cloud opacity \citep{fort07,fort08}.  The pressure-temperature profiles for these models are shown in Figure \ref{contrib}.  While profiles are shown in the right panel, we will first examine the left panel, which shows normalized contribution functions (\emph{cf}) for the thermal flux from the atmosphere \citep[e.g.][]{chamhunt87,griff98}:

\begin{equation}
cf(P)=B(\lambda,T)\frac{\mathrm{d}e^{-\tau}}{\mathrm{d}\log(P)}
\end{equation}

The band-averaged \emph{cf}s show the fractional contribution of various pressures to the outgoing thermal radiation in the chosen infrared bands.  These were computed by calculating the \emph{cf}s at 2000 wavelengths across the planet's spectrum, and then, at every pressure, integrating these \emph{cf}s across the IR bandpasses.  Importantly, the peak in the \emph{cf} differs from the pressure one would estimate from simply solving for the point where the actual temperature equals the brightness temperature \citep{chamhunt87}.  The \emph{cf}s show that there is considerable overlap in the contributions of the flux between the various \emph{Spitzer} bands.  While the peak in 24~\micron~emission is from pressures $2-3$ times greater than at 8~\micron, the overlap is considerable.  In general the shape of the \emph{cf}s is the same on both hemispheres, but deviations can be seen in the 3.6 and 8.0~\micron~bands, which are affected by the increased methane abundance on the night side.  Most prominently at 8.0~\micron, the enhanced gaseous opacity in the upper atmosphere leads to a significant flux contribution from lower pressure regions.  If one does not see as deeply on the night side compared to the day side, this complicates the interpretation of temperature homogenization.

The \emph{cf} at 24~\micron~peaks at lower pressures than the 8~\micron~\emph{cf} as a result of the increased water absorption at longer wavelengths \citep[e.g.][]{fort07,burr08}.  Figure \ref{contrib} shows that for both the day and night sides, CO is favored over CH$_4$, but the CH$_4$ abundance is not negligible, and absorption bands from this molecule can be seen in model spectra.  The simple night-side profile, that of an isolated object at $T_{\rm eff}$=1500 K, yields a night-side synthetic spectrum that matches the night-side photometry, as we will show in Figure \ref{model_spectrum}.  However, this high $T_{\rm eff}$ model  must vastly over-predict the temperature of the deeper atmosphere, just below the IRAC bands' \emph{cf}s.  The actual night-side $T_{\rm eff}$ must be considerably lower than 1500 K, to avoid an energy budget problem for the planet.  To further illustrate the need for the planet to be relatively cool at higher pressures (as suggested by Barman 2008), we plot a 1D planet-wide average profile in red in Figure \ref{contrib}.  Much of the upper atmosphere of both the day and night profiles is warmer than this 1D planet-wide average model---this cannot be true at all pressures.  The drawbacks of simple day/night models computed in this fashion have been discussed previously \citep{burr06,burr08} and argue for realistic 3D simulations of energy transport and atmospheric temperature structures.  In light of these uncertainties our computed \emph{cf}s should be regarded with care, but as we show below, these same atmosphere models are a good fit for the mid-infrared photometric data, which argues for the veracity of the treatment of chemistry and opacity.

The issue of the methane abundance on the day and night hemispheres is an important one.  \citet{swain08} recently reported a detection of methane absorption in the transmission spectrum of HD~189733b, which probes the region around the day-night terminator.  It is reasonable to expect that the cooler night side might have a higher methane abundance than the day side, but non-equilibrium carbon chemistry could alter this balance.  \citet{coop06} find mixing time scales faster than the chemical conversion time scale of CO to methane in hot Jupiter atmospheres, which leads to a day/night homogenization of the methane and CO abundances.  At this point it is too early to make definite statements regarding this issue for HD~189733b.  Observations of the planet's day-side emission spectrum around the 2.2~\micron~methane~band as well as the computation of spectra from 3D dynamical models \citep{fort06,show08} would both help to constrain the methane abundance on the day- and night-side hemispheres.

If an optically thick cloud deck exists at altitudes above the photosphere expected from purely gaseous opacity, both 8 and 24~\micron~radiation could emanate from the same pressure (that of the cloud top) and hence sense the same temperature structure between day and night.  However, a high opaque cloud of silicates or iron, which are the most likely candidates, is not expected from equilibrium chemistry at these modest temperatures.  Observations of the planet's optical transmission spectrum \citep{pont08,red08} suggest that it may have a haze layer that is reducing the depth of the observed absorption features \citep[for a more detailed discussion of this effect see][]{fort05}.  However, the detection of water and methane absorption features in the planet's near-IR ($1.4-2.5$~\micron) transmission spectrum indicates that this haze must be composed of relatively small particles, and it is extremely unlikely that it would affect the planet's emission spectrum at the relatively long ($>3$~\micron) wavelengths of our observations \citep{tinn07,swain08}.  A thick cloud layer would also lead to a featureless infrared emission spectrum, thus the detection of absorption features in the planet's $3.6-24$~\micron~broadband day-side emission spectrum \citep{char08,bar08} provides a somewhat weaker constraint on the presence of an opaque cloud layer above the upper range of the 8 and 24~\micron~contribution functions.  

If the 8 and 24~\micron~photosphere pressures indeed differ, then our observations constrain the rate at which the temperature changes with height in this region of the atmosphere.  The difference between the maximum and minimum hemisphere-averaged 8 and 24~\micron~brightness temperatures is $38\pm 48$~K for the maximum temperature and $27\pm70$~K for the minimum temperature.  This suggests a structure where the temperature varies only weakly with pressure.  In contrast, over a factor of two increase in pressure, a convective adiabat changes temperature by an amount $\Delta T_{\rm ad} = 0.7 R T/c_p$, where $R$, $T$, and $c_p$ are the specific gas constant, temperature, and specific heat at constant pressure.  For HD 189733b, where $R\approx 3700\,{\rm J}\,{\rm kg}^{-1}\,{\rm K}^{-1}$, $T\approx 1000\,$K, and $c_p\approx 1.3\times10^4\,{\rm J}\,{\rm kg}^{-1}\, {\rm K}^{-1}$, this expression yields $\Delta T_{\rm ad}\approx 200\,$K.  Thus, our observations suggest that the temperature increases with depth more weakly than an adiabat and hence that the atmosphere is not convective at these altitudes.  This result is consistent with the pressure-temperature profiles plotted in Figure \ref{contrib} and the predictions of previous 1D radiative-equilibrium calculations \citep[e.g.][]{fort05,seag05,burr05,bar05}. Three-dimensional circulation models produce similar predictions, indicating that temperature should increase with pressure on the night side but remain close to isothermal or exhibit an inversion layer on the day side \citep{coop05,coop06,show08,dobb08}.

\subsection{Spatially Resolved Atmospheric Features}

Our 24~\micron~flux maps (\S\ref{phase_fits} and Figure \ref{map}) indicate that the highest-flux region lies eastward of the substellar point, providing further evidence for the horizontal and/or vertical advection of the temperature field by jet streams, waves, or other processes.  The eastward phase shift of the high-flux region inferred here, $20-30$\degr~of longitude, is robust to model assumptions (2, 3, 4, or 6 slices) and is furthermore consistent with our previous inferences at 8~\micron~(Figure \ref{map}), which detected an eastward offset of $\sim30^{\circ}$ of longitude.  Our models also indicate that the lowest-flux region lies eastward of the antistellar point, although the size of this eastward shift varies substantially with model assumptions (ranging from $30^{\circ}$ longitude for a 2-slice map to $90^{\circ}$ longitude for a 4-slice map), due to the short duration of the data available before the transit.

At first glance the 8 and 24~\micron~maps in Figure \ref{map} appear to give conflicting answers for the location of the cold region on the planet's night side, but the lack of data before the transit at 8~\micron~means that we cannot rule out the presence of a second, larger cold spot centered $\sim$90\degr~west of the substellar point, similar to the one hinted at in the 24~\micron~data.  The error bars in the 12-slice model fit plotted in Fig. 3 of Paper I indicate that the flux for the slice located 90\degr~west of the substellar point is only $1\sigma$ higher than the lowest-flux slice, which is located 150\degr~east of the substellar point.  Without the benefit of the additional data before the transit available to us at 24~\micron, we cannot determine if the minimum observed in the planet's integrated 8~\micron~phase curve is a local minimum or a global minimum.  This idea is supported by the new phase-shifted four-slice fit to these data plotted in Figure \ref{map}, which shows that the 8~\micron~data can be fit consistently by a model with two minima, one located to the west of the substellar point and one located to the east.

Several circulation models predict that the regions of maximum and minimum flux will be shifted to the east of the substellar and antistellar points, respectively, by amounts analogous to that suggested by our flux maps \citep{show02,coop05,fort06,show08}.  These models do not reproduce the cold spot east of the substellar point indicated by the 8~\micron~data (this places it in the same hemisphere as the hot region near the substellar point, which is also shifted to the east), but our new 24~\micron~data argue for a global minimum located to the west of the substellar point, which would be consistent with these models.  Other models suggest that the flow may contain hot and cold vortices that migrate in longitude; depending on timing, this could cause eastward or westward offsets of the flux minima and maxima \citep{cho03,cho08,rau07,rau08}.  If we are to obtain strong constraints on the location of minima and other features occurring in the region to the west of the substellar point, it will require additional observations spanning the missing half of this planet's phase curve.

\subsection{Matching Models to Observations}

\begin{figure*}
\epsscale{0.8}
\plotone{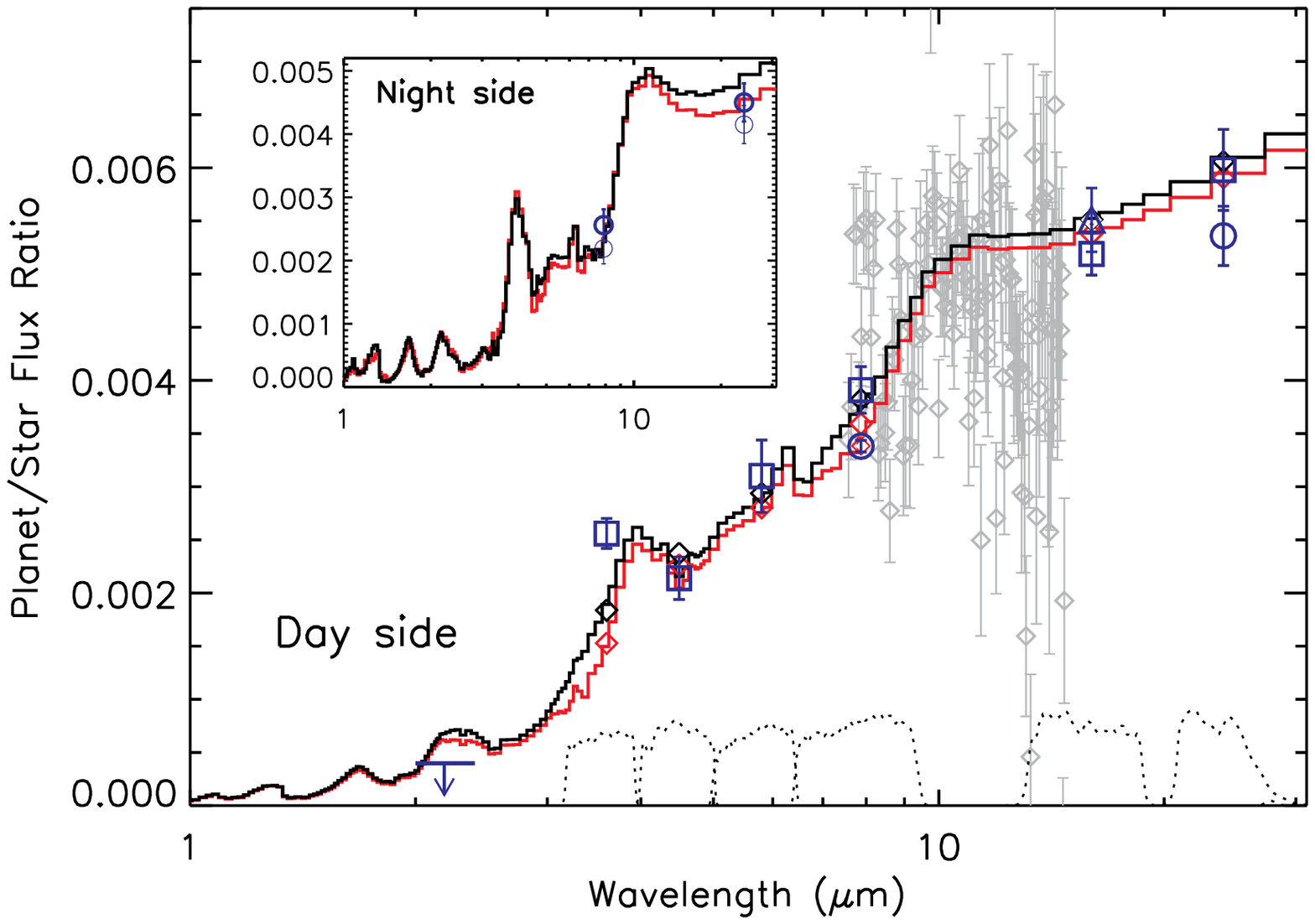}
\caption{Planet-to-star flux ratios as a function of wavelength.  Photometric data are in blue and error bars are 1$\sigma$.  The data from \citet{char08} are squares, the IRS 16~\micron~value from \citet{dem06} is a triangle, and our data from Paper I and this paper are circles.  The line at 2.2~\micron~is the upper limit from \citet{barn07}.    In light gray is the \citet{grill07} IRS spectrum ratio.  The transmission functions for the \emph{Spitzer} bandpasses are shown as dotted curves at the bottom of the figure.  Model spectral ratios are shown in black and red. Diamonds show the values of these models averaged over the \emph{Spitzer} bandpasses.  The red model (with a fixed methane mixing ratio of $7\times 10^{-6}$, see text) fits the data well except for the 3.6~\micron~point, as does the black model, which uses equilibrium chemistry mixing ratios.   Inset:  On the night side, thin circles are the uncorrected flux ratios while the thick circles show the flux ratios after accounting for the effects of star spots.  As on the day side, models with a fixed methane mixing ratio and equilibrium mixing ratios both fit the data well.
\label{model_spectrum}}
\end{figure*}

In light of the mid-infrared day-side emission spectrum \citep{char08} and near infrared transmission spectrum \citep{swain08} that have recently become available for this planet, it is worthwhile to take a global view of HD~189733b and attempt to reconcile current atmosphere models with our estimates for the day- and night-side fluxes as well as these other data.  In Figure \ref{model_spectrum} we show ratio spectra, computed from the profiles shown in Figure \ref{contrib}, as updated from \citet{fort07}.  For the planet's day side, we show a model that utilizes equilibrium chemical abundances (black), and in red a model that uses a non-equilibrium CH$_4$/CO ratio of 0.014 (CH$_4$ mixing ratio of $7\times 10^{-6}$) as taken directly from \citet{coop06} and \citet{fort06}.  This particular CH$_4$ abundance is consistent with the upper limit on the CH$_4$ abundance on the planet's limb of $5\times 10^{-5}$ derived by \citet{swain08}.  Both day side models assume the atmosphere absorbs incident flux and redistributes this energy evenly over the day side, with no energy transported to the night side, as this provides the best fit to the relatively high mid-IR fluxes observed on the day side (blue points).  It is immediately apparent that both day-side models are an excellent (1$\sigma$) fit to the data, with the exception of the IRAC 3.6 $\mu$m point and the \citet{barn07} K-band 1$\sigma$ upper limit.

We model the night side as an object in isolation and vary the effective temperature in order to match the observed fluxes at 8 and 24~\micron.  \citet{burr06} and \citet{bar08} have both investigated similar kinds of models.  In this case we find that a model with $T_{\rm eff} = 1500$ K provides the best fit, although this model has an implausibly hot interior for a planet.  \citet{bar08} find very similar fits using the 8~\micron~data from Paper I alone, and propose that the high night-side fluxes might be explained by the transport of energy from the day side via flows \emph{below the mid IR photosphere} probed by \emph{Spitzer}.  As long as such flows remained below the level of the  mid-IR photosphere, day-side flux originating from higher up in the atmosphere would appear to be consistent with a no-recirculation model, while the night side would have increased emission resulting from this added influx of energy.   

As suggested by \citet{bar08}, increased circulation below the level of the mid-IR photosphere would also help to explain the depressed day-side $K$-band flux, as the effective photosphere in this bandpass is below that probed by most mid-infrared wavelengths (see Figure \ref{contrib}).  In this scenario the day-side pressure-temperature profile would be considerably colder than that predicted from a simple 1D model for pressures greater than $\sim$0.1 bar. However, we note that emission from the $K$-band and the IRAC 3.6~\micron~band emerges from similar pressure ranges, such that a depression in flux in $K$-band but an enhancement in the IRAC 3.6~\micron~band would be difficult to explain with this model.  Perhaps abundant day-side methane could be suppressing the $K$-band flux; this bandpass spans a strong methane absorption band located around 2.3~\micron.  \citet{fort06a} modeled this planet's atmosphere and showed that a reasonable methane abundance would produce observable effects in the planet's $K$-band emission.  The recent detection of methane absorption in the planet's transmission spectrum \citep{swain08} confirms that this molecule is indeed present at the day-night terminator, although these observations do not provide a strong constraint on the methane abundances on the day and night hemispheres.

If we are to reconcile fully the planet's day- and night-side emission spectra, it will require more sophisticated 1D models than we have yet computed, or better yet, full 3D simulations with non-gray radiative transfer, which we are working towards.  Such models would go a long way towards characterizing both the planet's energy budget and the manner and depth dependence of temperature homogenization in its atmosphere.  Based on the simpler models and observations described above, we concur with \citet{bar08}, who suggested that 43\% of the energy absorbed on the day side must be emitted on the night side.  In Barman's framework the revised night-side flux ratios we present in this paper, which take into account the effects of star spots, argue for even more efficient redistribution.  Detections of the planet's flux at near-infrared wavelengths (such as $J$, $H$, or $K$) which sample the peak of the planet's emission spectrum and probe higher atmospheric pressures than those viewed by \emph{Spitzer}, are required in order to constrain the redistribution efficiency in a less model-dependent manner.  

\section{Conclusions}

There are several clear conclusions that emerge from these observations.  The planet's atmosphere exhibits only a modest variation in the day/night brightness temperatures at 8 and 24~\micron~when compared to radiative-equilibrium predictions for highly-irradiated, tidally-locked planets.  This implies efficient transport of thermal energy from the day side to the night side by atmospheric winds at the level of both the 8 and 24~\micron~photospheres.  The planet's 8 and 24~\micron~phase curves both reach a peak before the secondary eclipse, indicating that the hottest region on the day side is shifted $20-30$\degr~east of the substellar point at the location of both photospheres and providing additional evidence for the horizontal and/or vertical advection of the temperature field in the planet's atmosphere.  The similarities between the phase curves at 8~and 24~\micron~suggests that either both wavelengths sense similar atmospheric pressures or that the circulation behaves in a fundamentally similar fashion across the relatively modest (factor of $2-3$) range in pressures that atmosphere models indicate are sensed here.  The uncertain abundance of methane in the planet's atmosphere complicates this picture \citep{swain08}, but based on other data we can definitively rule out the presence of an opaque cloud layer above the range of the 8 and 24~\micron~contribution functions that, if present, might affect these observations \citep{char08,bar08}.

Our observations at 24~\micron~confirm that HD~189733b's phase curve is fundamentally different than that of $\upsilon$~Andromedae b or HD 179949b.  Although these three planets constitute a very limited sample, this result would seem to point towards the existence of two distinct classes of hot Jupiter atmospheres, characterized by either efficient or inefficient thermal homogenization between the day and night sides of the planet.  This is particularly interesting in light of recent results by \citet{knut08} and \citet{char08}, who characterized the day-side broadband emission spectra for both HD~189733b and  HD~209458b.  These observations revealed that HD~209458b has an atmospheric temperature inversion with water bands in emission instead of absorption \citep{burr07}, while HD~189733b's spectrum is well-described by a model with no temperature inversion and water absorption bands \citep{bar08}.  Although this is also a very limited sample, these results point towards a similar division of hot Jupiters into two distinct classes.

It is possible that the presence of a temperature inversion and the degree of thermal homogenization may be connected.  \citet{burr07} and \citet{fort08}, following up on prescient earlier work by \citet{hub03}, have both suggested that HD 209458b's temperature inversion might be caused by gas-phase TiO/VO, which would have condensed out of HD~189733b's cooler atmosphere.  In this picture, temperature inversions would be correlated with the degree of irradiation, with a distinct division between the two classes of planets set by the condensation temperature of TiO/VO \citep{fort08,burr08}.  \citet{fort08} also point out that planets with temperature inversions absorb more of the incident flux higher in their atmospheres, where the radiative time scale is short compared to the advective time scale.  They argue that this would naturally lead to large day-night temperature differences for these planets, while planets with lower levels of irradiation and no temperature inversions would be more homogenized.  We note that both $\upsilon$~Andromedae b and HD 179949 have higher levels of incident flux than HD~189733b, placing them in the same class as HD~209458b.  Unfortunately neither of these planets are eclipsing, making it difficult to check directly for the presence of a temperature inversion.  A better test would be to measure the day-night temperature difference for HD~209458b, which clearly does have a temperature inversion, or HD 149026b, whose high 8~\micron~brightness temperature \citep{har07} strongly favors the presence of a temperature inversion \citep{fort06a,burr08}.  We have obtained such observations of HD 209458b, and will report on the results in a future paper.

Although HD 209458b has a temperature inversion while HD 189733b does not, they still share a number of basic characteristics, including a gas-dominated structure with minimal to no solid core.  It is not at all clear that information on the atmospheric dynamics of these planets will be applicable to smaller, core-dominated planets such as HD~149026b and GJ~436b \citep{butler04,sato05,fort06a,gill07,dem07,torr07,adam08,winn08}.  These planets likely have atmospheres enriched in heavy elements, perhaps by a factor of 10 or more.  In the solar system, there is a clear correlation between the percentage of planet mass that is core, and atmospheric metallicity \citep{lod03}.  Uranus and Neptune have a C/H ratio of 30-40 times solar while Jupiter's is only ~3 times solar.  The higher surface gravities and potentially differing atmospheric compositions of HD~149026b and GJ~436b may significantly alter the nature of the circulation between their presumably tidally-locked day and night sides.  Moreover, GJ~436b is likely only pseudo-synchronized, as it has an orbital eccentricity of 0.15 \citep{dem07,demory07}, which further complicates this picture.  Secondary eclipse observations may shed some light on the properties of these planets, but they provide only a snapshot of the global properties of the day-side atmosphere.  Observations of the phase variations of these two planets would provide a considerably richer source of information on their spatially-resolved properties; such information is crucial if we are to understand the nature of atmospheric circulation for this distinctly different class of planets.  

Circulation models for all of these planets would also benefit from observations at additional wavelengths.  Our 1D radiative-equilibrium calculations indicate that the 3.6 and 24~\micron~bandpasses should span the widest possible range of pressures for HD~189733b, from $0.001-1$ bars.  The same may hold true for HD~209458b, but its higher day-side temperatures should lead to an increased mid infrared opacity \citep{fort08}, which may shift the \emph{cf}s to lower pressures.  The 4.5 and 5.8~\micron~band contribution functions overlap considerably with the 8 and 24~\micron~bands for HD~189733b, so it is possible that little new information would be obtained from observations in these bandpasses.  However, the 3.6~\micron~band (as well as the near-infrared $J$, $H$, and $K$ bands) probes deeper atmospheric pressures that are closer to the peak in this planet's spectral energy distribution.  For IRAC specifically, the 3.6 and 4.5~\micron~bands are particularly sensitive to absorption by CO, methane, and water, and since they lie closer to the maximum in the planet's flux, give a more robust measure of the day/night \emph{effective} temperatures.  Although the intrapixel sensitivity evident in both detectors \citep{reach05,char05,mor06,knut08} presents a challenge for observations of phase variations in these two bandpasses, this effect is increasingly well-understood and it should be possible to develop a robust correction with a modest investment of additional \emph{Spitzer} time.  We recommend that observations of other planets in these bandpasses span entire planetary orbits if possible; this would provide an additional check on the increasingly large effects of star spots at these shorter wavelengths, as well as resolving any ambiguities in the longitudinal temperature distributions of these planets.  

Such time-intensive observations in the IRAC 3.6 and 4.5~\micron~bandpasses would be well-matched to the proposed non-cryogenic \emph{Spitzer} mission.  Although \emph{Spitzer} is predicted to run out of cryogen in spring 2009, observations in the two shortest-wavelength channels should continue to achieve the same sensitivity after the cryogen is exhausted.  \emph{Spitzer} is the only continuing observatory that currently offers the means to study the infrared phase curves of extrasolar planets.  Beyond \emph{Spitzer}, we must await the launch of JWST, as such observations would be incredibly challenging from the ground.

\acknowledgments

We would like to thank David Latham, Matthew Holman, Joshua Winn, Gilbert Esquerdo, Jos\'e Fernandez, Gaspar Bakos, and C\'esar Fuentes for their assistance in obtaining observations of HD~189733~with the FLWO 1.2 m telescope, and for sharing time on their previously scheduled nights to obtain these observations.  We thank Mark Marley for helpful discussions, and the referee for a thoughtful and detailed review.  This work is based on observations made with the \emph{Spitzer Space Telescope}, which is operated by the Jet Propulsion Laboratory, California Institude of Technology, under contract to NASA.  We also utilize observations made with KeplerCam, which was developed with partial support from the Kepler mission under Cooperative Agreement NCC2-1390. Support for this work was provided by NASA through an award issued by JPL/Caltech.  HAK was supported by a National Science Foundation Graduate Research Fellowship.

\end{document}